\documentclass[aps,amssymb,amsmath,pra,reprint]{revtex4-2}

\usepackage{graphicx}
\usepackage{dcolumn}
\usepackage{bm}

\begin{document}


 \title{Excitation of wakefields in carbon nanotubes: a hydrodynamic model approach}
\author{P. Mart\'in-Luna}\email{pablo.martin@uv.es}
\affiliation{Instituto de F\'isica Corpuscular (IFIC), Universitat de Val\`encia - Consejo Superior de Investigaciones Cient\'ificas, 46980 Paterna, Spain}
\author{A. Bonatto}
\affiliation{Graduate Program in Information Technology and Healthcare Management, and the Beam Physics Group,
Federal University of Health Sciences of Porto Alegre, Porto Alegre, RS, 90050-170, Brazil}
\author{C. Bontoiu}
\affiliation{
Department of Physics, The University of Liverpool, Liverpool L69 3BX, United Kingdom, \\ The Cockcroft Institute, Sci-Tech Daresbury, Warrington WA4 4AD, United Kingdom}
\author{G. Xia}
\affiliation{
Department of Physics and Astronomy, The University of Manchester, Manchester M13 9PL, United Kingdom \\ 
The Cockcroft Institute, Sci-Tech Daresbury, Warrington WA4 4AD, United Kingdom}
\author{J. Resta-L\'opez}\email{javier2.resta@uv.es} 
\affiliation{Instituto de Ciencia de los Materiales (ICMUV), Universidad de Valencia, 46071 Valencia, Spain}

\date{\today}

\begin{abstract}
The interactions of charged particles with carbon nanotubes may excite electromagnetic modes in the electron gas produced in the cylindrical graphene shell constituting the nanotube wall. This wake effect has recently been proposed as a potential novel method of short-wavelength high-gradient particle acceleration. In this work, the excitation of these wakefields is studied by means of the linearized hydrodynamic model. In this model, the electronic excitations on the nanotube surface are described treating the electron gas as a 2D plasma with additional contributions to the fluid momentum equation from specific solid-state properties of the gas. General expressions are derived for the excited longitudinal and transverse wakefields. Numerical results are obtained for a charged particle moving within a carbon nanotube, paraxially to its axis, showing how the wakefield is affected by parameters such as the particle velocity and its radial position, the nanotube radius, and a friction factor, which can be used as a phenomenological parameter to describe effects from the ionic lattice. Assuming a particle driver propagating on axis at a given velocity, optimal parameters were obtained to maximize the longitudinal wakefield amplitude.


\end{abstract}

\maketitle


\section{Introduction}\label{Introduction}

Carbon nanotubes (CNTs) were discovered by S. Iijima in 1991 \cite{iijima1991CNTdiscovery} and they can be thought of as a sheet of graphene (a hexagonal lattice of carbon) rolled into a cylinder. CNTs can exhibit metallic or semiconductor properties depending on their rolling pattern (i.e. on their radius and geometrical angle). Thus, as a consequence of their unique thermo-mechanical and electronic properties and dimensional flexibility, CNTs have been widely studied in both theoretical and experimental aspects. As a hollow structure, CNTs may be used for transporting and focusing charged particles similar to crystal channeling. In particular, experimental results on 2 MeV $\text{He}^{+}$ ions \cite{Zhu2005_SPIE_channel_ion} and 300 keV electrons \cite{Chai2007_channel300kev_electrons} channeling in CNTs have been reported. 

On the other hand, solid-state wakefield acceleration using crystals was proposed in the 1980s and 1990s by T. Tajima and others \cite{TajimaCavenago1987_XrayAccelerator_PhysRevLett.59.1440, chen1987solid, chen1997crystal} as an alternative particle acceleration technique to sustain TV/m acceleration gradients. In the original Tajima’s conceptual scheme \cite{TajimaCavenago1987_XrayAccelerator_PhysRevLett.59.1440}, a longitudinal electric wakefield is excited by a laser (laser driven) in the crystal so that a properly injected witness beam may be accelerated. Similarly, the ultrashort charged particle bunches (beam driven) can excite electric wakefields so that the energy loss of the driving bunch can be transformed into an increment of energy of
a witness bunch. However, the angstrom-size channels of natural crystals pose a limitation for the beam intensity acceptance and the dechanneling rate. In this context, CNTs can obtain wider channels in two dimensions and longer dechanneling lengths \cite{BIRYUKOV2002_PLB_channelCNTs, BELLUCCI2005_PLB_channelCNT}, which together with their remarkable electronic properties, larger degree of dimensional flexibility and thermo-mechanical strength, make them a robust candidate for TeV/m acceleration. Consequently, carbon nanostructures (CNTs or even graphene layers) are currently being widely studied for wakefield acceleration \cite{Bonatto2023_effective_plasma_POP, Bontoiu2023_catapult}.

Wakefields in CNTs are excited through the collective oscillation of electrons on the nanotube's surface, often referred to as plasmons. This excitation arises from the interaction between the driving bunch and the CNT, leading to the generation of these wakefields. The electronic excitations on the single-wall CNT surface produced by the interaction with charged particles have been theoretically studied using a dielectric theory \cite{Arista2001IonsPhysRevA.64.032901, Arista2001ChargedParticlesPhysRevB.63.165401, WangMiskovic2002_dielectric_theory_PhysRevA.66.042904}, a hydrodynamic model \cite{Stockli2001PhysRevB.64.115424, WangMiskovic2004_Hydro_theory_PhysRevA.69.022901, MOWBRAY2004_hydrodynamic_model_wake_effects}, a two-fluid model \cite{WangMiskovic2004TwoFluidModelPhysRevB.70.195418, Nejati2009_doi:10.1063/1.3077306}, a quantum hydrodynamic model \cite{WeiWang2007_quantum_hydrodynamic_model_PhysRevB.75.193407} and a kinetic model \cite{Song2006_KineticModel_PhysRevA.78.012901, Zhao2008_ChinPhysLett_KineticModel, Zhang2013_WakeEffects_ChinPhysLett_KineticModel, ZHANG2014_kinetic_model_Carbon}. However, while these articles mostly explore properties such as the energy loss and stopping power or, at most, evaluate the induced surface electron density and/or induced potential \cite{MOWBRAY2004_hydrodynamic_model_wake_effects, Zhang2013_WakeEffects_ChinPhysLett_KineticModel, ZHANG2014_kinetic_model_Carbon}, they do not address the induced longitudinal and transverse wakefields, which could provide acceleration and focusing, respectively, for a witness charge.


Thus, this article is motivated by the need to study the wakefields excited by charged particles moving paraxially inside CNTs. Additionally, the article aims to optimize the parameters of the CNT in order to obtain the highest longitudinal wakefield for particle acceleration purposes.

This work is organized as follows. In Sec. \ref{Linearized hydrodynamic theory} the general expressions are derived for the longitudinal and transverse wakefields excited by the interaction of a charged particle with a CNT in the realm of the hydrodynamic model. This model has been chosen because of its simplicity and its good agreement with the dielectric formalism in random-phase approximation \cite{WangMiskovic2004_Hydro_theory_PhysRevA.69.022901}. In Sec. \ref{Results and discussion}, after investigating the influence of different model parameters in terms of dispersion relations, the CNT parameters are optimized to achieve the highest longitudinal wakefield for a given driving velocity. Finally, the main conclusions of this study are presented in Sec. \ref{Conclusions}.

\section{Linearized hydrodynamic theory}\label{Linearized hydrodynamic theory}
\begin{figure}[!h]
\includegraphics[width=0.8\columnwidth]{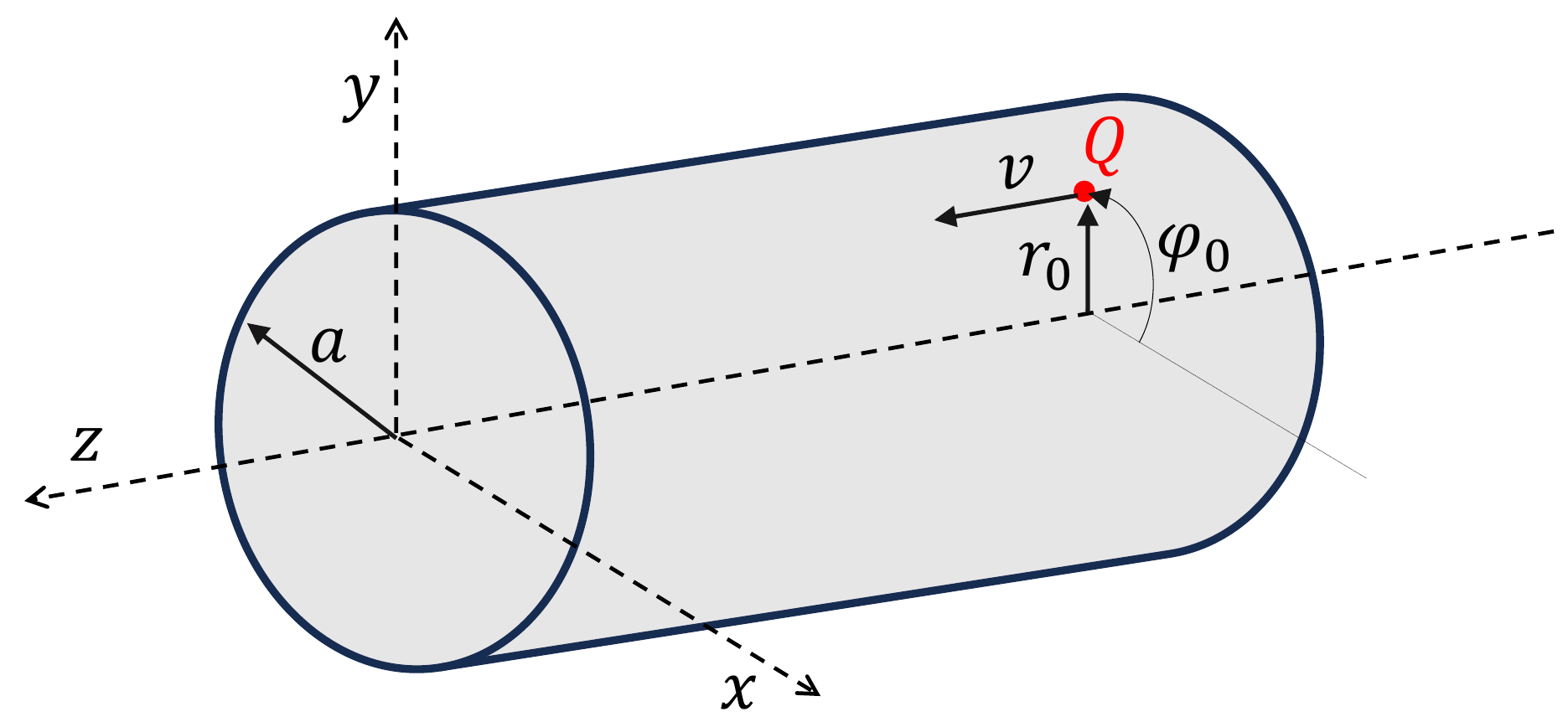}
\centering
\caption{Scheme of the considered charge $Q$ travelling parallel to the $z$-axis inside a tube.}
\label{fig:CNT_scheme}
\end{figure}

In this work, a linearized hydrodynamic theory \cite{Stockli2001PhysRevB.64.115424, WangMiskovic2004_Hydro_theory_PhysRevA.69.022901, Nejati2009_doi:10.1063/1.3077306} is adopted, in particular a model which includes single-electron based excitations on nanotube surfaces and was described by Wang and Mi\v{s}kovic in \cite{WangMiskovic2004_Hydro_theory_PhysRevA.69.022901}, although we will use the SI units instead of atomic units. In this theory, a single-wall CNT is modelled as an infinitesimally thin and infinitely long cylindrical shell with a radius $a$. The delocalized electrons of the carbon ions are considered as a two-dimensional free-electron gas that is confined over the cylindrical surface of the CNT with a uniform surface density $n_0$. It is considered a driving point-like charge $Q$ travelling parallel to the $z$-axis inside the tube with a constant velocity $v$ (see Fig. \ref{fig:CNT_scheme}). Consequently, its position as a function of time $t$ is $\mathbf{r}_0(t)=\left(r_0, \varphi_0, vt\right)$ in cylindrical coordinates. As a consequence of the presence of the driving charge $Q$, the homogeneous electron gas will be perturbed and can be modelled as a charged fluid with a velocity field $\mathbf{u}(\mathbf{r}_a ,t)$ and a perturbed density per unit area $n_1(\mathbf{r}_a ,t)$, where $\mathbf{r}_a=(a, \varphi,z)$ are the coordinates of a point at the cylindrical surface of the tube. As the electron gas is confined to the cylindrical surface, the normal component to the surface of the tube of the velocity field $\mathbf{u}$ is zero.

In the linearized hydrodynamic model, the electronic excitations on the tube wall can be described by three differential equations: (i) the continuity equation

\begin{equation}\label{continuity_equation}
\frac{\partial n_1\left(\mathbf{r}_a, t\right)}{\partial t}+n_0 \nabla_{\|} \cdot \mathbf{u}\left(\mathbf{r}_a, t\right)=0,
\end{equation}

\noindent (ii) the Poisson's equation 

\begin{equation}\label{Poisson_equation}
\nabla^2 \Phi(\mathbf{r}, t)=\frac{1}{\varepsilon_0}\left[e n_1\left(\mathbf{r}_a, t\right) \delta(r-a)-Q \delta\left(\mathbf{r}-\mathbf{r}_0\right)\right],
\end{equation}

\noindent and (iii) the momentum-balance equation

\begin{equation}\label{momentum_balance_equation}
\begin{split}
\frac{\partial \mathbf{u}\left(\mathbf{r}_a, t\right)}{\partial t}=&\frac{e}{m_e} \nabla_{\|} \Phi\left(\mathbf{r}_a, t\right)-\frac{\alpha}{n_0} \nabla_{\|} n_1\left(\mathbf{r}_a, t\right)+\\
&\frac{\beta}{n_0} \nabla_{\|}\left[\nabla_{\|}^2 n_1\left(\mathbf{r}_a,t\right)\right]-\gamma \mathbf{u}\left(\mathbf{r}_a, t\right).
\end{split}
\end{equation}

\noindent In these equations, $\mathbf{r}=(r,\varphi,z)$ is the position vector, $\nabla=\hat{\mathbf{r}}\frac{\partial}{\partial r} +\hat{\boldsymbol{\varphi}}\frac{1}{r} \frac{\partial}{\partial \varphi} +\hat{\mathbf{z}}\frac{\partial}{\partial z} $,  $\nabla_{\|}=\hat{\boldsymbol{\varphi}}\frac{1}{a} \frac{\partial}{\partial \varphi} +\hat{\mathbf{z}}\frac{\partial}{\partial z}$ differentiates only tangentially to the tube surface, $\Phi$ the electric scalar potential, $e$  the elementary charge, $m_e$ the rest mass of the electron, $\varepsilon_0$ the vacuum electric permittivity and $\delta$ the Dirac delta. Equation (\ref{momentum_balance_equation}) shows the sum of four different contributions. The first term in the right-hand side is the force on electrons on the nanotube surface due to the tangential component of the electric field generated by the driving charge $Q$ and the consequent perturbed density $n_1$. The second and third terms are related to the parts of the internal interaction force in the electron gas. In particular, the second term takes into account the possible coupling with acoustic modes defining the parameter $\alpha=v^2_F/2$ (in which $v_F=\hbar(2\pi n_0)^{1/2}/m_e$ is the Fermi velocity of the two-dimensional electron gas; $\hbar$ is the reduced Planck constant), and the third term is a quantum correction that arises from the functional derivative of the Von Weizsacker gradient correction in the equilibrium kinetic energy of the electron fluid \cite{Nejati2009_doi:10.1063/1.3077306} and describes single-electron excitations in the electron gas, where the parameter $\beta=\frac{1}{4}(\frac{\hbar}{m_e})^2$ has been defined. The last term is introduced to satisfy the non-conservation of the system and represents a frictional force on electrons due to scattering with the ionic-lattice charges, where $\gamma$ is the friction parameter. The friction parameter may be also used as a phenomenological parameter to take into account the broadening of the plasmon resonance in the excitation spectra of different materials \cite{Arista2001IonsPhysRevA.64.032901}. 

Taking into account that the electric potential vanishes at $r \rightarrow \infty$ and is finite at the origin $r=0$, the potential can be expanded in terms of the modified Bessel functions $I_m(x)$ and $K_m(x)$ of integer order $m$ (i.e. a Fourier-Bessel expansion). The total potential inside the nanotube ($r<a$) can be calculated as $\Phi_{in}=\Phi_0+\Phi_{ind}$ with the Coulomb potential $\Phi_0$ due to the driving charge and the induced potential $\Phi_{ind}$ due to the perturbation of the electron fluid on the CNT surface. The total potential outside the nanotube ($r>a$) will be denoted as $\Phi_{out}$. Thus, the Fourier-Bessel expansion of these three components is
\begin{equation}
\begin{split}
\Phi_0(r, \varphi, \zeta)=&\frac{1}{4 \pi \varepsilon_0} \frac{Q}{\left\|\mathbf{r}-\mathbf{r}_{\mathbf{0}}\right\|}=\frac{Q}{4 \pi^2 \varepsilon_0} \sum_{m=-\infty}^{+\infty} \int_{-\infty}^{+\infty} \mathrm{d} k \\
&e^{i k \zeta+i m\left(\varphi-\varphi_0\right)} I_m\left(|k| r_{\text{min}}\right) K_m\left(|k| r_{\text{max}}\right),
\end{split}
\end{equation}

\begin{equation}
\begin{split}
\Phi_{i n d}(r, \varphi, \zeta)=&\frac{Q}{4 \pi^2 \varepsilon_0} \sum_{m=-\infty}^{+\infty} \int_{-\infty}^{+\infty} \mathrm{d} k \\
&e^{i k \zeta+i m\left(\varphi-\varphi_0\right)} I_m\left(|k| r_0\right) I_m(|k| r) A_m(k),
\end{split}
\end{equation}

\begin{equation}
\begin{split}
\Phi_{out}(r, \varphi, \zeta)=&\frac{Q}{4 \pi^2 \varepsilon_0} \sum_{m=-\infty}^{+\infty} \int_{-\infty}^{+\infty} \mathrm{d} k \\
&e^{i k \zeta+i m\left(\varphi-\varphi_0\right)} I_m\left(|k| r_0\right) K_m(|k| r) B_m(k),
\end{split}
\end{equation}

\noindent where $k$ is the wavenumber, a comoving coordinate $\zeta=z-vt$ has been defined and $r_{\text{min}}=\text{min}(r,r_0)$, $r_{\text{max}}=\text{max}(r,r_0)$. The unknown coefficients $A_m(k)$ and $B_m(k)$ can be calculated if the following boundary conditions are imposed: (i) the continuity of the electric potential at the nanotube surface
\begin{equation}
\left.\Phi_{i n}(r, \varphi, \zeta)\right|_{r=a}=\left.\Phi_{o u t}(r, \varphi, \zeta)\right|_{r=a},
\end{equation}

\noindent and (ii) the discontinuity of the radial component of the electric field due to perturbed density $n_1$ of the electron fluid
\begin{equation}
\left.\frac{\partial \Phi_{o u t}(r, \varphi,\zeta)}{\partial r}\right|_{r=a}-\left.\frac{\partial \Phi_{i n}(r, \varphi, \zeta)}{\partial r}\right|_{r=a}=\frac{en_1(a,\varphi, \zeta)}{\varepsilon_0}.
\end{equation}

 \noindent Thus, the coefficients $A_m(k)$ and $B_m(k)$ are given by the following non-dimensional functions:
\begin{equation}
A_m(k)=\frac{\Omega_p^2 a^2\left(k^2+m^2 / a^2\right) K_m^2(|k| a)}{D_m(k)},
\end{equation}
\begin{equation}
B_m(k)=\frac{D_m(k)+\Omega_p^2 a^2\left(k^2+\frac{m^2}{a^2}\right) K_m(|k| a) I_m(|k| a)}{D_m(k)},
\end{equation}


\noindent where $\Omega_p=\sqrt{\frac{e^2 n_0}{\varepsilon_0 m_e a}}$ is the plasma frequency, 
\begin{equation}
D_m(k)=k v(k v+i \gamma)-\omega_m^2(k),
\end{equation}
and 

\begin{equation}\label{eq:wm}
\begin{split}
\omega_m^2(k)=&\alpha\left(k^2+\frac{m^2}{a^2}\right)+\beta\left(k^2+\frac{m^2}{a^2}\right)^2+ \\
&\Omega_p^2 a^2\left(k^2+\frac{m^2}{a^2}\right) K_m(|k| a) I_m(|k| a).
\end{split}
\end{equation}

\noindent Consequently, the resonant excitations occur when $kv=\omega_m(k)$ if the damping or friction factor $\gamma$ vanishes.

On the other hand, the longitudinal and transverse electric wakefields inside the tube are, respectively,
\begin{equation}\label{eq:Wz}
\begin{aligned}
W_z(r, \varphi, \zeta) &=-\frac{\partial \Phi_{i n}}{\partial z} =\frac{Q}{4 \pi^2 \varepsilon_0} \sum_{m=-\infty}^{+\infty} \int_{-\infty}^{+\infty} \mathrm{d} k\, k e^{i m\left(\varphi-\varphi_0\right)} \\
&\times \left(I_m\left(|k| r_{\text{min}}\right) K_m\left(|k| r_{\text{max}}\right) \sin (k \zeta)\right. \\
& +I_m\left(|k| r_0\right) I_m(|k| r)\left[\text{Re}\left[A_m(k)\right] \sin (k \zeta)\right. \\
& \left.\left.+\text{Im}\left[A_m(k)\right] \cos (k \zeta)\right]\right)=W_{z0}+W_{z1}+W_{z2},
\end{aligned}
\end{equation}

\begin{equation}\label{eq:Wr}
\begin{aligned}
W_r(r, \varphi, \zeta) &=-\frac{\partial \Phi_{i n}}{\partial r} =-\frac{\partial \Phi_{0}}{\partial r}-\frac{Q}{4 \pi^2 \varepsilon_0} \sum_{m=-\infty}^{+\infty} \int_{-\infty}^{+\infty} \mathrm{d} k \\
&|k| e^{i m\left(\varphi-\varphi_0\right)} 
I_m\left(|k| r_0\right) I_m^{\prime}(|k| r)\\
&\times\left[\text{Re}\left[A_m(k)\right] \cos (k \zeta)\right.
-\text{Im}\left[A_m(k)\right] \sin (k \zeta)] \\
&=W_{r0}+W_{r1}+W_{r2},
\end{aligned}
\end{equation}
\noindent where $I^{\prime}_m(x)=dI_m(x)/dx$ and the properties $ \text{Re}\left[A_m(k)\right]=\text{Re}\left[A_m(-k)\right]$ and $\text{Im}\left[A_m(k)\right]=-\text{Im}\left[A_m(-k)\right]$ were considered; $\text{Re}$ and $\text{Im}$ denote the real and imaginary part, respectively. The previous integrals have been separated in three different terms: the first terms $W_{z0}$ and $W_{r0}$ come from the Coulomb potential and the other terms $W_{z1}, W_{z2}$ and $W_{r1}, W_{r2}$ from the induced potential. It is important to note that the third terms $W_{z2}$ and $W_{r2}$ can be analytically integrated if the damping factor vanishes ($\gamma \rightarrow 0^+$): 

\begin{equation}\label{eq:Wz2}
\begin{aligned}
&W_{z2}(r, \varphi, \zeta)=\frac{Q}{4 \pi^2 \varepsilon_0} \sum_{m=-\infty}^{+\infty} \int_{-\infty}^{+\infty} \mathrm{d} k\, k e^{i m\left(\varphi-\varphi_0\right)} \\
&I_m\left(|k| r_0\right) I_m(|k| r)\text{Im}\left[A_m(k)\right] \cos (k \zeta) \\
&=\frac{-Q}{2 \pi \varepsilon_0} \sum_{m=-\infty}^{+\infty} e^{i m\left(\varphi-\varphi_0\right)} k_m I_m\left(k_m r_0\right) I_m\left(k_m r\right) \\
&\Omega_p^2 a^2\left(k_m^2+\frac{m^2}{a^2}\right) K_m^2\left(k_m a\right)\left|\frac{\partial Z_m}{\partial k}\right|_{k=k_m}^{-1} \cos \left(k_m \zeta\right),
\end{aligned}
\end{equation}

\begin{equation}\label{eq:Wr2}
\begin{aligned}
&W_{r2}(r, \varphi, \zeta)=\frac{Q}{4 \pi^2 \varepsilon_0} \sum_{m=-\infty}^{+\infty} \int_{-\infty}^{+\infty} \mathrm{d} k\, |k| e^{i m\left(\varphi-\varphi_0\right)} \\
&I_m\left(|k| r_0\right) I_m^{\prime}(|k| r)\text{Im}\left[A_m(k)\right] \sin (k \zeta) \\
&=\frac{-Q}{2 \pi \varepsilon_0} \sum_{m=-\infty}^{+\infty} e^{i m\left(\varphi-\varphi_0\right)} k_m I_m\left(k_m r_0\right)I_m^{\prime}\left(k_m r\right) \\
&\Omega_p^2 a^2\left(k_m^2+\frac{m^2}{a^2}\right) K_m^2\left(k_m a\right)\left|\frac{\partial Z_m}{\partial k}\right|_{k=k_m}^{-1} \sin \left(k_m \zeta\right),
\end{aligned}
\end{equation}

\noindent where it has been defined the quantity $Z_m(k)=\text{Re}[D_m(k)]=(k v)^2-\omega_m^2(k)$ and $k_m$ are the (positive) roots of $Z_m(k)$, i.e. the condition of the plasma resonance $k_mv=\omega_m(k_m)$.

\section{Results and discussion} \label{Results and discussion}

As it can be deduced from Eqs. (\ref{eq:Wz2})-(\ref{eq:Wr2}), the roots $k_m$ given by the plasma resonance are essential to describe the behaviour of the wakefields. For this reason, this section begins with a detailed analysis of the dispersion relation.

\subsection{Dispersion relation}\label{Dispersion relation}
\begin{figure}[!h]
\includegraphics[width=0.93\columnwidth]{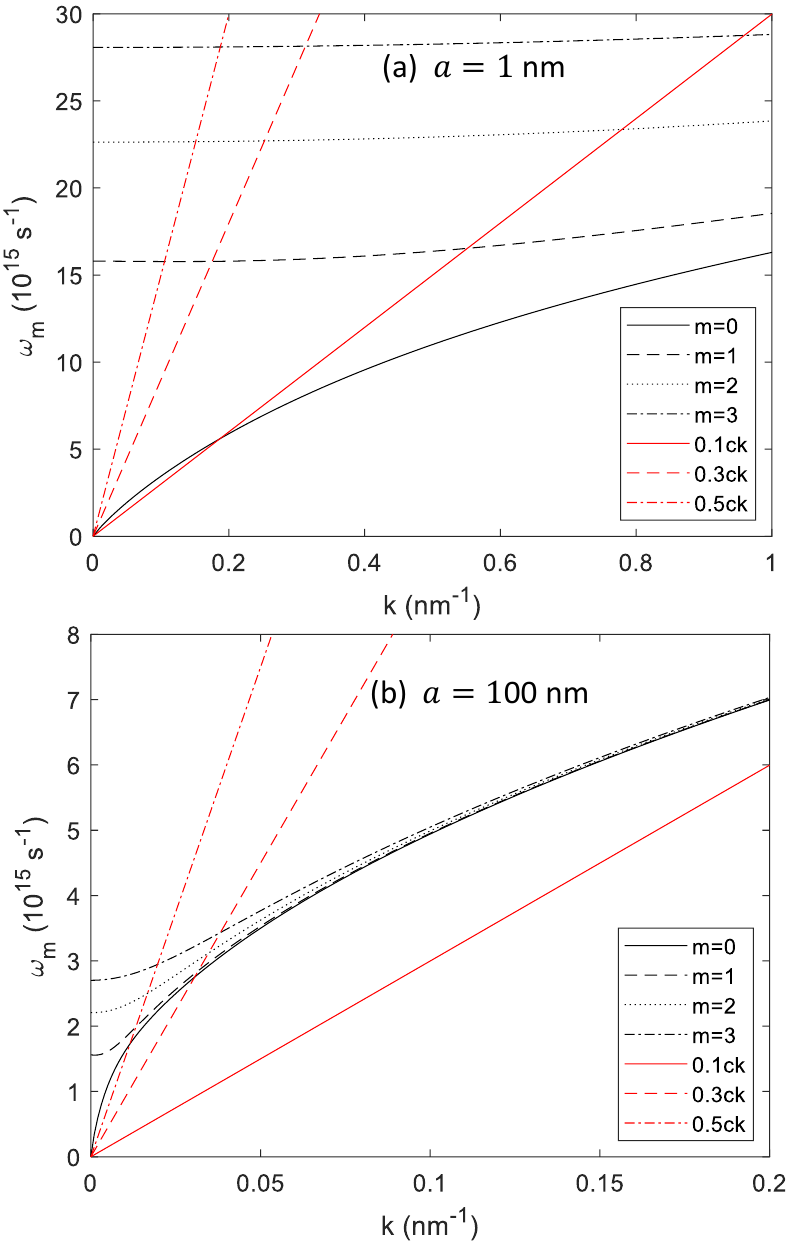}
\caption{Dispersion curves $\omega_m(k)$ for several angular-momentum modes for (a) $a=1$\,nm and (b) $a=100$\,nm. The resonances $k_m$ are the intersection of the $kv$ lines (plotted for $v=0.1c, v=0.3c$ and $v=0.5c$) with the dispersion curves $\omega_m(k)$.}
\label{fig: Dispersion_modes}
\end{figure}

In the following calculations, unless otherwise indicated, it is assumed that the surface electron density of a single-wall carbon nanotube can be approximated by the electron-gas density of a graphite sheet: $n_0=1.53\times10^{20}$\,m$^{-2}$ \cite{Ostling1997_surface_density_PhysRevB.55.13980, WangMiskovic2004_Hydro_theory_PhysRevA.69.022901}. Figure \ref{fig: Dispersion_modes} shows the dispersion curves $\omega_m(k)$ for the first modes at different CNT radii. If the radius is too small, the first mode $m=0$ does not satisfy the resonance condition for high velocities, while modes with $m>0$ have a solution $k_m$ for those velocities, as seen in Fig. \ref{fig: Dispersion_modes}(a) at $a=1$\,nm. If the CNT radius increases, the resonance condition can be satisfied for a wider range of high velocities, as it is depicted in Fig. \ref{fig: Dispersion_modes}(b) for $a=100$\,nm. Furthermore, it can be seen that the modes $\omega_m(k)$ converge for a sufficiently large value of the wavenumber $k$. Therefore, if the resonance condition $k_m$ is sufficiently large, then all modes will have a similar value of $k_m$. Note that a large value of $k_m$ indicates that the associated wavelength of the wakefield ($\lambda_m=2\pi/k_m$) will be smaller. Moreover, if the surface density $n_0$ increases, then the dispersion curves increase and the resonance conditions are obtained for higher $k_m$.

\begin{figure}[!h]
\includegraphics[width=0.93\columnwidth]{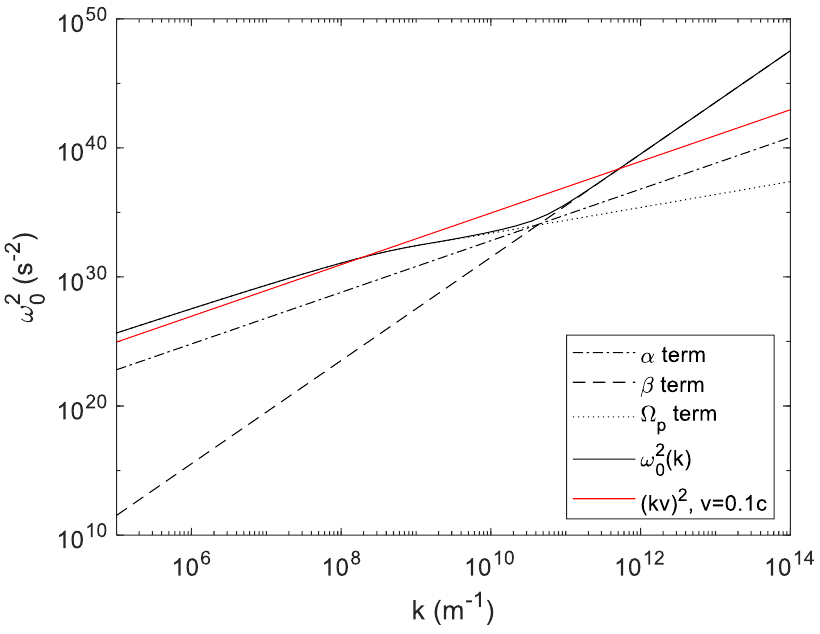}
\caption{Dispersion relation $\omega_0^2(k)$ and contribution from the different three terms of Eq. (\ref{eq:wm}) for $a=1$\,nm.}
\label{fig: Dispersion_terms}
\end{figure}

\begin{figure*}[t!]
\includegraphics[width=2\columnwidth]{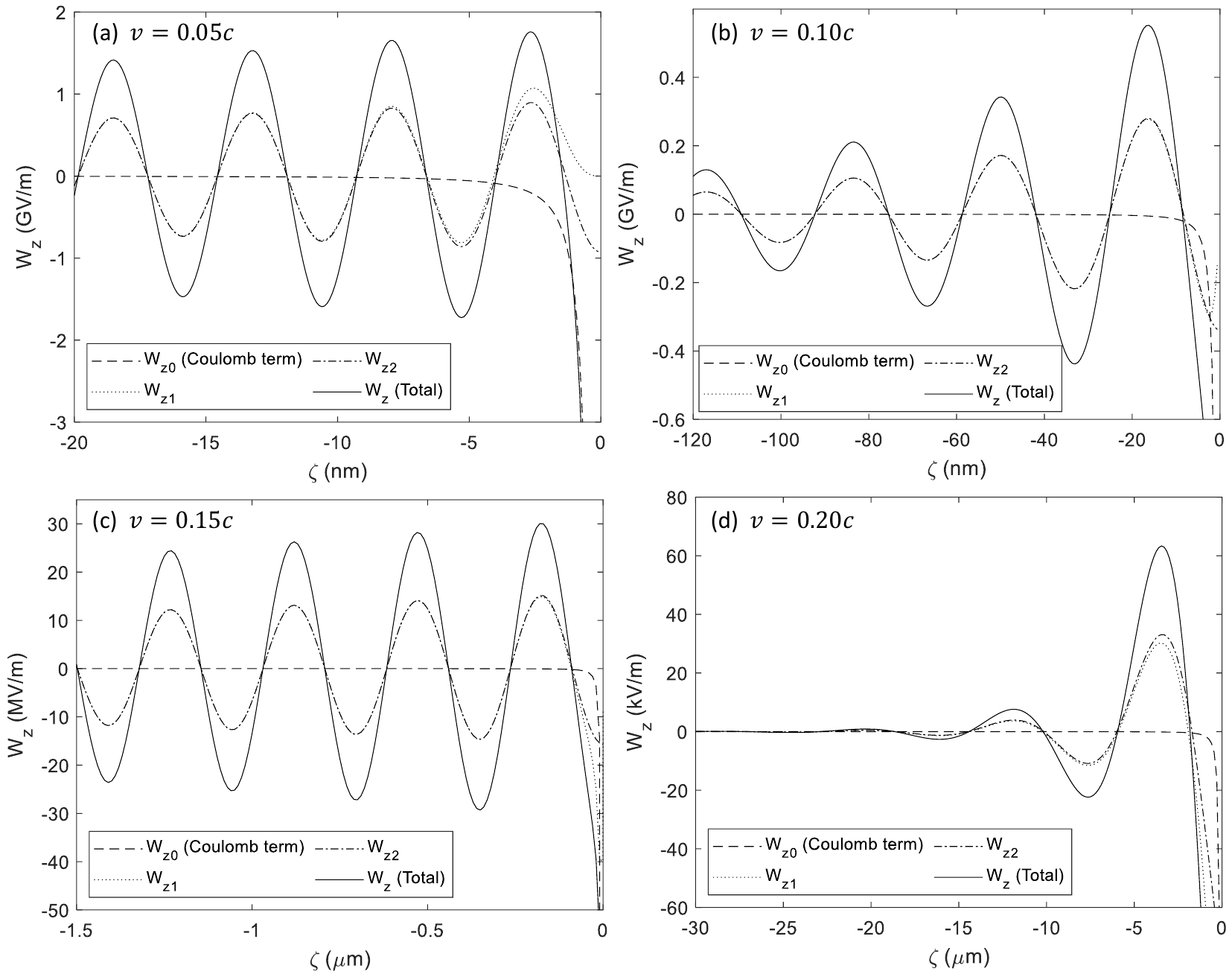}
\caption{Longitudinal wakefield contributions on axis ($r=0$) for a proton travelling on axis ($r_0=0$) at different velocity values: (a) $v=0.05c$, (b) $v=0.10c$, (c) $v=0.15c$ and (d) $v=0.20c$. The CNT radius is $a=1$\,nm and the friction parameter is $\gamma=0.01\Omega_p$ in cases (a) and (b), and $\gamma=0.0001\Omega_p$ in (c) and (d). Note that the driving proton is at $\zeta=0$.}
\label{fig: Wakefield_terms}
\end{figure*}

Furthermore, it is analysed the fundamental mode $\omega_0(k)$, since it is the only mode which contributes for a particle travelling on axis (or if the wakefield is calculated on axis). Figure \ref{fig: Dispersion_terms} depicts the dependence of the fundamental mode $\omega_0(k)$ of the resonant frequency on the wavenumber $k$ for a CNT with a radius $a=1$\,nm as well as the contribution of the three addends in Eq. (\ref{eq:wm}). It can be observed that the contribution from the $\Omega_p$ term dominates for low values of $k$, whereas the contribution from the $\beta$ term is dominant for larger values of $k$.
Nevertheless, there is an intermediate region, $10^{10}\,\text{m}^{-1}\lesssim k\lesssim 10^{11}\,\text{m}^{-1}$, where the three contributions exhibit considerable similarity. The resonance condition $k_m$ for a given velocity $v$ is given by the intercept of $(kv)^2$ (a parallel line to the $\alpha$ contribution) with $\omega_m^2(k)$. Hence, the resonance condition cannot be satisfied if $v<\alpha$, where $\alpha=v_F/\sqrt{2}$. In general, two resonances $k_m$ can exist (as seen in Fig. \ref{fig: Dispersion_terms} for $v=0.1c$) and then the contribution from both resonances must be summed in Eqs. (\ref{eq:Wz2})-(\ref{eq:Wr2}). However, the contribution from the resonance with a larger value of $k_m$ is, in general, totally negligible taking into account the exponential behaviour of the Bessel function $K_m(x)=\sqrt{\frac{\pi}{2x}}e^{-x}$ for $x\rightarrow \infty$ and the factor $\left|\frac{\partial Z_m}{\partial k}\right|_{k=k_m}^{-1}$ that decreases rapidly for high values of $k_m$. Therefore, by resonance condition we practically mean the first root $k_m$ (the resonances shown in Fig. \ref{fig: Dispersion_modes}). Consequently, for high velocities $v$, the resonance condition $k_m$ is obtained (if it is satisfied) in the region where the $\Omega_p$ term dominates and contribution from the parameters $\alpha$ and $\beta$ is negligible. Thus, the parameters $\alpha$ and $\beta$ only become important if the resonance condition is obtained in the intermediate region.

\subsection{Electric wakefields}\label{Electric wakefields}

\begin{figure}[!h]
\includegraphics[width=\columnwidth]{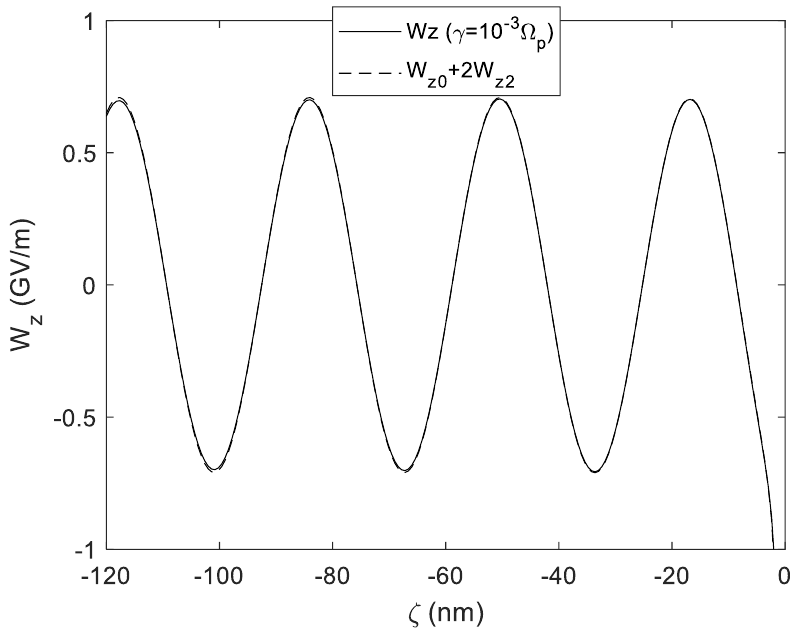}
\caption{Comparison between Eq. (\ref{eq:Wz}) for small $\gamma$ and the approximation using Eq. (\ref{eq:Wz2}) to calculate $W_{z2}$ for $a=1$\,nm and $v=0.1c$.}
\label{fig:comparison_gamma}
\end{figure}

In the following calculations, it is considered that the point-like charged particle is a proton, i.e. $Q=e$. Figure~\ref{fig: Wakefield_terms} shows the three different contributions to the longitudinal wakefield (cf. Eq. (\ref{eq:Wz})) for different driving velocities. It can be seen that the Coulomb term is only important near the driving particle, whereas $W_{z1}$ and $W_{z2}$ are responsible for the plasmonic excitations and are practically identical, except in the proximity of the driving particle. Moreover, the wavelength of the wakefield increases with the velocity $v$ as it was deduced from the dispersion relation. Besides, the wakefield amplitude decreases as the proton speed increases since we are approaching to the velocities that do not satisfy the resonance condition. The friction parameter $\gamma$ produces an exponential decay of the wakefield ($W_{z1}+W_{z2}$) with the distance behind the driving charged particle. For this reason, the value of $\gamma$ has been diminished in Fig.~\ref{fig: Wakefield_terms}(c)-(d) in order to see the plasmonic excitations. Thus, when the friction parameter $\gamma$ is very small, $W_{z2}$ follows a cosine pattern and the longitudinal wakefield can be approximated by $W_z=W_{z0}+2W_{z2}$ ($W_{z2}$ calculated using Eq. (\ref{eq:Wz2})), as shown in Fig.~\ref{fig:comparison_gamma}.


\begin{figure}[!h]
\includegraphics[width=\columnwidth]{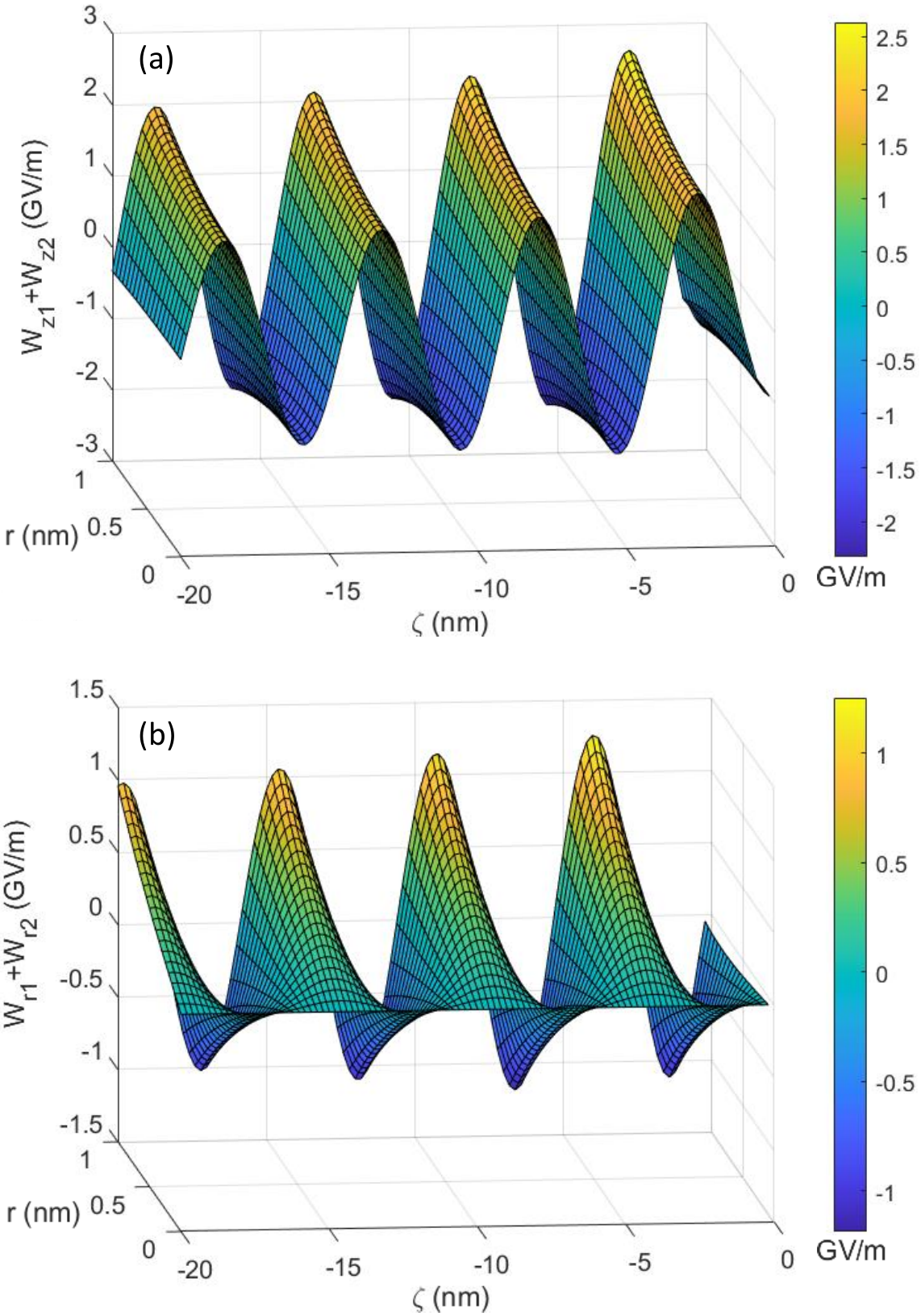}
\caption{(a) Longitudinal and (b) transverse components of the plasmonic wakefields inside a CNT considering the following parameters: $a=1$\,nm, $v=0.05c$ and $\gamma=0.01\Omega_p$.}
\label{fig:wakefields_inside_tube}
\end{figure}

Figure \ref{fig:wakefields_inside_tube} depicts an example of the longitudinal and transverse wakefield inside a CNT. Here, to better appreciate the wakefield details, only the contribution from the plasmonic excitation is shown, taking into account that the Coulomb contribution is negligible except near the driving particle. In Fig. \ref{fig:wakefields_inside_tube} one remarkable thing is that plasmonic wakefields increase with the radial distance $r$ because of the dependence on $I_m(|k|r)$ and $I_m^\prime(|k|r)$ (both increasing functions with the argument) of $W_z$ and $W_r$, respectively. A similar reasoning can be used to show that the wakefields are higher if the particle travels off axis ($r_0\neq0$), although in this case higher order modes ($|m|>0$) should be computed. These effects become more important for lower velocities, since the value of the resonance wavenumber increases.

\begin{figure}[!h]
\includegraphics[width=\columnwidth]{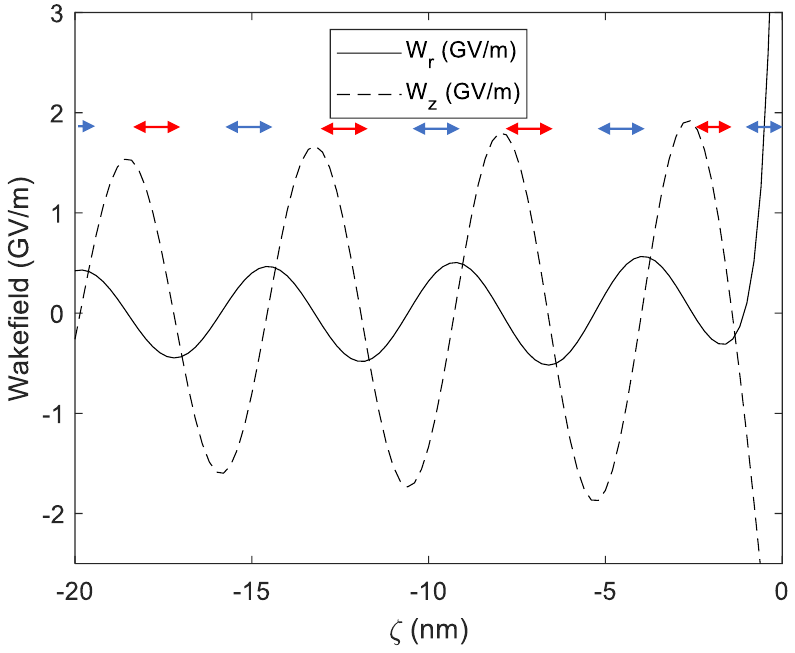}
\caption{Wakefields (including the Coulomb term) at $r=a/2$ for $a=1$\,nm, $v=0.05c$, $\gamma=0.01\Omega_p$ and $r_0=0$. The red (blue) arrows indicate the regions where a positive (negative) witness charged particle would experience both acceleration and focusing simultaneously.}
\label{fig:wakefields_at_r05a}
\end{figure}

Furthermore, it is interesting to note that there is a phase offset of $\pi/2$ between the longitudinal and transverse wakefields. These results agree with the Panofsky-Wenzel theorem \cite{PanofskyWenzel1956} and are similar to what is observed for wakefields excited in homogeneous plasmas in the linear regime \cite{esarey2009physics_RevModPhys.81.1229}. As a consequence, there are periodical regions where the witness charged particles can simultaneously experience both acceleration and focusing (if they travel off-axis), as it is depicted in Fig. \ref{fig:wakefields_at_r05a}.


\subsection{Optimization of the CNT parameters}\label{Optimization of the CNT parameters}

\begin{figure}[!h]
\includegraphics[width=\columnwidth]{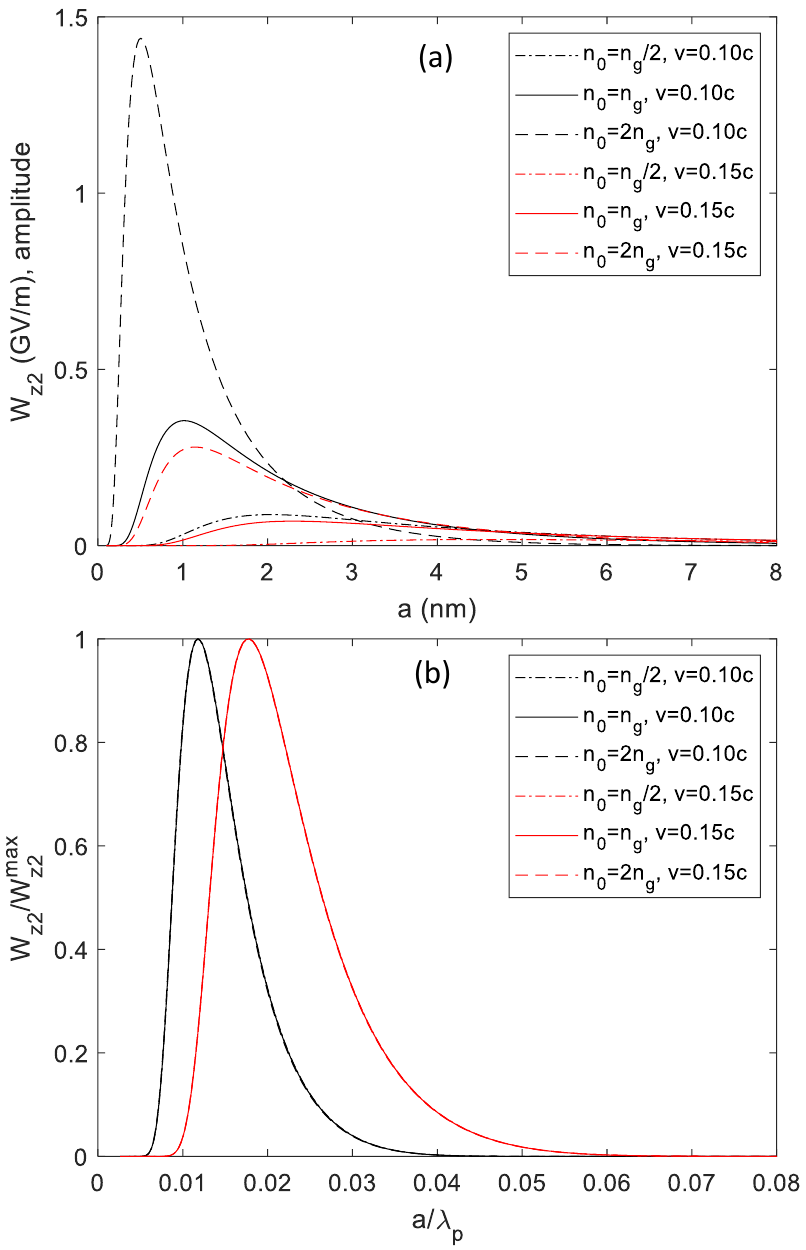}
\caption{(a) Amplitude of $W_{z2}$ as a function of the radius $a$  for different values of $n_0$ and $v$ ($n_g=1.53\times10^{20}$\,m$^{-2}$ is the electron-gas density of a graphite sheet). (b) Amplitude of $W_{z2}$ (normalized to the maximum of Fig. \ref{fig:Wz2_vs_a}(a)) as a function of $a/\lambda_p$.}
\label{fig:Wz2_vs_a}
\end{figure}

\begin{figure}[!h]
\includegraphics[width=\columnwidth]{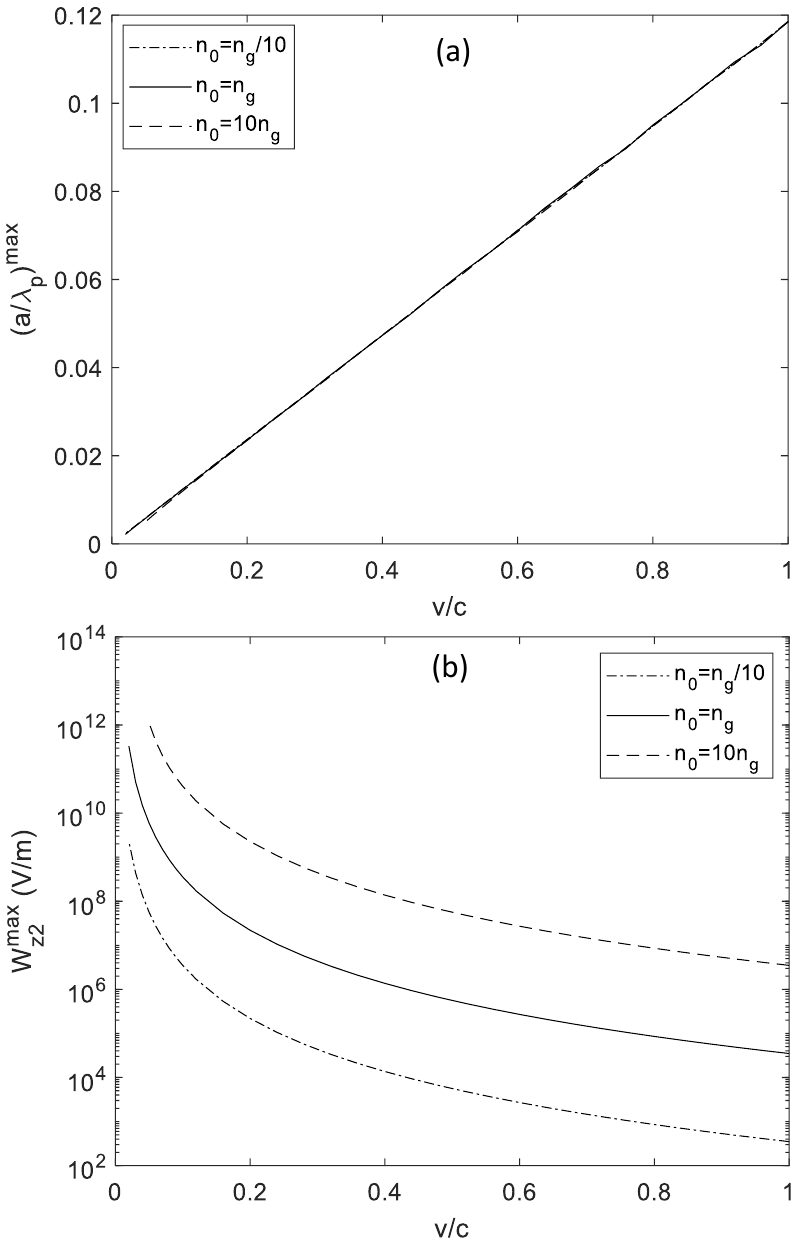}
\caption{(a) Optimum radius and (b) maximum longitudinal wakefield associated as a function of $v$ for different surface densities ($n_g=1.53\times10^{20}$\,m$^{-2}$ is the electron-gas density of a graphite sheet).}
\label{fig:optimum_parameters_vs_v}
\end{figure}

As pointed out in the previous section, the plasmonic excitations can be approximated by $2W_{z2}$ when the friction parameter $\gamma$ converges to zero in the considered system.  Therefore, Eq. (\ref{eq:Wz2}) can be used to efficiently optimize key parameters, such as $n_0$, $a$ and $v$, to enhance the longitudinal wakefield amplitude. Concretely, this section is focused on the plasmonic excitation created on axis by a proton travelling on axis, i.e. $r=r_0=0$. Figure \ref{fig:Wz2_vs_a}(a) depicts the amplitude of $W_{z2}$ as a function of the radius $a$. The maximum of the wakefield increases with the surface density $n_0$ and moves to smaller radii $a$. It is also worth noting that these plots do not depend on the surface density if both the radius and the wakefield amplitude are normalized to the plasma wavelength $\lambda_p=2\pi c/\Omega_p$ and the peak maximum $W_{z2}^{max}$, respectively (see Fig. \ref{fig:Wz2_vs_a}(b)). Thus, there is an optimum radius (in units of $\lambda_p$) for a given driving velocity regardless of the surface density $n_0$. This optimum radius $(a/\lambda_p)^{max}$ is proportional to the driving velocity $v$, as it can be seen in Fig. \ref{fig:optimum_parameters_vs_v}(a). On the other hand, Fig. \ref{fig:optimum_parameters_vs_v}(b) shows the peak maximum $W_{z2}^{max}$ as a function of $v$ for different surface densities. The maximum wakefield is obtained for lower velocities (as long as they satisfy the resonance condition) and increases with the surface density. Thus, driving particles with low velocities can excite more efficiently plasmonic modes in CNTs. However, at very low velocities, the resonance condition cannot be satisfied. As a result, the curves in Fig. \ref{fig:optimum_parameters_vs_v}(b) do not originate at $v=0$.

In summary, for a given velocity, the maximum longitudinal wakefield is obtained for the optimum CNT radius given by Fig. \ref{fig:optimum_parameters_vs_v}(a) and a surface density as high as possible. Finally, it is important to remark that for high velocities we should take into account the relativistic effects which are not considered in this manuscript. Nevertheless, it is worth mentioning that a behavior qualitatively similar to that shown in Fig. \ref{fig:Wz2_vs_a}(b) was also observed in Fig. 4 of \cite{Bonatto2023_effective_plasma_POP}, where particle-in-cell (PIC) simulations were used with a driving bunch composed of 1 GeV electrons (with an energy spread of 1$\%$). In that case, the maximum wakefield was obtained for $a/\lambda_p\approx0.10c$, a value quite similar to what was obtained in Fig. \ref{fig:optimum_parameters_vs_v}(a) as $v\rightarrow c$. Therefore, the CNT radius optimization carried out in this manuscript might provide a good approximation of its value.


\section{Conclusions}\label{Conclusions}

The linearized hydrodynamic model in conjunction with the Poisson's equation has been used to study the electric wakefields generated by a point-like charge travelling parallel to the axis in a CNT. General expressions have been derived for the longitudinal and transverse wakefields and their dependencies on the surface density, the CNT radius and the velocity of the driving charged particle have been numerically studied and related to the dispersion relation. It has been shown that the friction parameter produces an exponential decay of the excited plasmonic modes. If the friction is negligible, the plasmonic excitations can be approximated by twice the Eqs. (\ref{eq:Wz2})-(\ref{eq:Wr2}). This approximation for the longitudinal wakefield was used to perform an optimization of the CNT radius (in units of the plasma wavelength) for a given driving velocity. Interestingly, at least at a qualitative level, the results agree with those obtained in \cite{Bonatto2023_effective_plasma_POP} through PIC models. Hence, the linearized hydrodynamic model might be used to obtain an approximation of the optimum radius without requiring time-consuming PIC simulations. In future works, a quantitative comparison will be performed.

It is worth mentioning that the wakefields generated by a point-like charge could be used as a Green’s function to compute the wakefields excited by a driving bunch with an arbitrary charge distribution. It has been shown that a single proton may excite GV/m wakefields in CNTs with nanometric radii. Consequently, a beam with $10^6$ protons would be able to excite PV/m wakefields, provided that the beam size is small enough to simply consider the sum of the contribution of each single particle. However, the obtention
of beams with such small sizes can prove to be extremely challenging. Nevertheless, in the case of ultra-relativistic driving charged particles, the optimal radius and wavelength associated with peak wakefields are estimated to be $\gtrsim 100$\,nm. Besides, space-charge effects are mitigated in beams with highly relativistic velocities. As a consequence, it would be easier to obtain beams which can be propagated inside the CNT and witness beams which fit in the periodic regions where they can experience acceleration and focusing for ultra-relativistic driving velocities. Even in this case, a driving beam with $10^6$ protons may excite TV/m wakefields, which are more than four orders of magnitude higher than those obtained with conventional RF cavities. Hence, the use of nano-structures may open new possibilities to obtain ultra-high particle acceleration gradients.

\begin{acknowledgments}
This work has been supported by Ministerio de Universidades (Gobierno de Espa\~{n}a) under grant agreement FPU20/04958, and the Generalitat Valenciana under grant agreement CIDEGENT/2019/058. 
\end{acknowledgments}

\providecommand{\noopsort}[1]{}\providecommand{\singleletter}[1]{#1}%

\end{document}